\documentclass[aps,showpacs,superscriptaddress,twocolumn,amsmath,amssymb,pra]{revtex4-1}
\usepackage{dcolumn}% Align table columns on decimal point
\usepackage{bm}% bold math
\usepackage[dvips]{graphicx}
\usepackage{wrapfig,subfigure}
\usepackage{amssymb}
\usepackage{color}
\usepackage{ulem}
\newcommand{\n}{\mbox{\boldmath $\nabla$}}

\newcommand{\Xb}{{\bf X}}

\newcommand{\ep}{\varepsilon}

\newcommand{\ad}{a^{\dagger}}

\begin{document}

\title{Quantum heating of a parametrically modulated oscillator: spectral signatures}

\author{M. I. Dykman}
\affiliation{Department of Physics and Astronomy, Michigan State University, East Lansing, Michigan 48824}
\author{M. Marthaler}
\affiliation{Institut f\"ur Theoretische Festk\"orperphysik
and DFG-Center for Functional Nanostructures (CFN), Karlsruhe Institute of Technology, D-76128 Karlsruhe, Germany}
\author{V. Peano}
\affiliation{Department of Physics and Astronomy, Michigan State University, East Lansing, Michigan 48824}
\affiliation{Freiburg Institute for Advanced Studies (FRIAS),
Albert-Ludwigs-Universit\"at Freiburg,
79104 Freiburg, Germany}

\begin{abstract}
We show that the noise spectrum of a parametrically excited nonlinear oscillator can display a fine structure. It emerges from the interplay of the nonequidistance of the oscillator quasienergy levels and quantum heating that accompanies relaxation. The heating leads to a finite-width distribution over the quasienergy, or Floquet states even for zero temperature of the thermal reservoir coupled to the oscillator. The fine structure is due to transitions from different quasienergy levels, and thus it provides a sensitive tool for studying the distribution. For larger damping, where the fine structure is smeared out, quantum heating can be detected from the characteristic double-peak structure of the spectrum, which results from transitions accompanied by the increase or decrease of the quasienergy.
\end{abstract}
\date{\today}


\maketitle

\section{Introduction}

Nonlinearity is advantageous for observing quantum effects in vibrational systems. It makes the energy levels nonequidistant and the frequencies of different inter-level transitions different, which in turn enables spectroscopic observation of the quantum energy levels. In addition, nonlinearity leads to an interesting behavior of vibrational systems in external periodic fields, including the onset of bistability of forced vibrations.  The interest in quantum effects in modulated nonlinear oscillators significantly increased recently due to the development of high-quality microwave resonators with the anharmonicity provided by Josephson junctions and to applications of these systems in quantum information \cite{Wallraff2004,Siddiqi2005,Watanabe2009,Mallet2009,Vijay2009,Wilson2010}.  The long-sought \cite{Blencowe2004a,*Schwab2005a} quantum regime has been reached also in nanomechanical systems \cite{O'Connell2010,Teufel2010,Riviere2010}. This development has opened the possibility of measurements of a single quantum nonlinear oscillator rather than of an ensemble of oscillators. In addition, the systems are  versatile and allow accessing different dynamical regimes.

One of the important problems that can be addressed with strongly modulated nonlinear oscillators is quantum fluctuations far from thermal equilibrium. In addition to the standard quantum uncertainty, such fluctuations come from the coupling of a quantum system to a thermal bath. The coupling leads to relaxation of the system via emission of excitations in the bath (photons, phonons, etc) accompanied by transitions between the system energy levels. If the coupling is weak, the transition rates are small compared to the transferred energy. In the classical case, the transitions lead to friction.

At the quantum level, however, one should take into account that the transitions happen at random. The randomness gives rise to a peculiar quantum noise and the related quantum heating of the oscilator. Its important manifestation is quantum activation in driven nonlinear oscillators \cite{Dykman1988a,*Dykman2007,Marthaler2006}, where the noise leads to activation-type transitions between the states of forced vibrations. Quantum activation has been now seen in the experiment \cite{Vijay2009}. There is also an observation of the quantum heating in the spectrum of a resonantly driven oscillator \footnote{F.~R.~Ong {\textit et al.}, in preparation (experiment) and M. Boissonneault {\textit et al.}, in preparation (theory); we are grateful to P. Bertet for informing us about this work}. However, to the best of our knowledge, no direct measurements of the relaxation-induced distribution over quantum states have been made and no direct means to measure this distribution have been proposed.

In this paper we show that the distribution over the states of a modulated nonlinear oscillator can be measured spectroscopically. If the oscillator is strongly underdamped, the power spectrum of its fluctuations and the spectrum of the response to an additional weak field can display a fine structure. The intensities of the fine-structure lines are directly related to the occupation of the oscillator states, and the line shapes depend on the  effective quantum temperature as well as the temperature of the thermal reservoir. We note that spectroscopy has been long recognized as a means of getting an insight into the dynamics of a strongly driven oscillator and, more recently, of using the oscillator for quantum measurements \cite{Dykman1979a,*Dykman1994b,Drummond1980c,*Drummond1981,Stambaugh2006a,*Chan2006,Nation2008,Serban2010,Vierheilig2010,Boissonneault2009,*Boissonneault2010,Laflamme2010}. However, the fine structure of the spectra has not been discussed earlier.

We will study the fine structure for an oscillator parametrically modulated at frequency $\omega_F$ close to twice the eigenfrequency $\omega_0$. Classically, as a result of parametric resonance the modulated oscillator can start vibrating at frequency $\omega_F/2$. The steady vibrational states are determined by the balance between the modulation, the dissipation due to coupling to a thermal bath, and the oscillator nonlinearity \cite{LL_Mechanics2004}. For not too strong modulation, the oscillator has two vibrational states, which have the same amplitude and differ in phase by $\pi$.

\begin{figure}[h]
\includegraphics[width=8.5 cm]{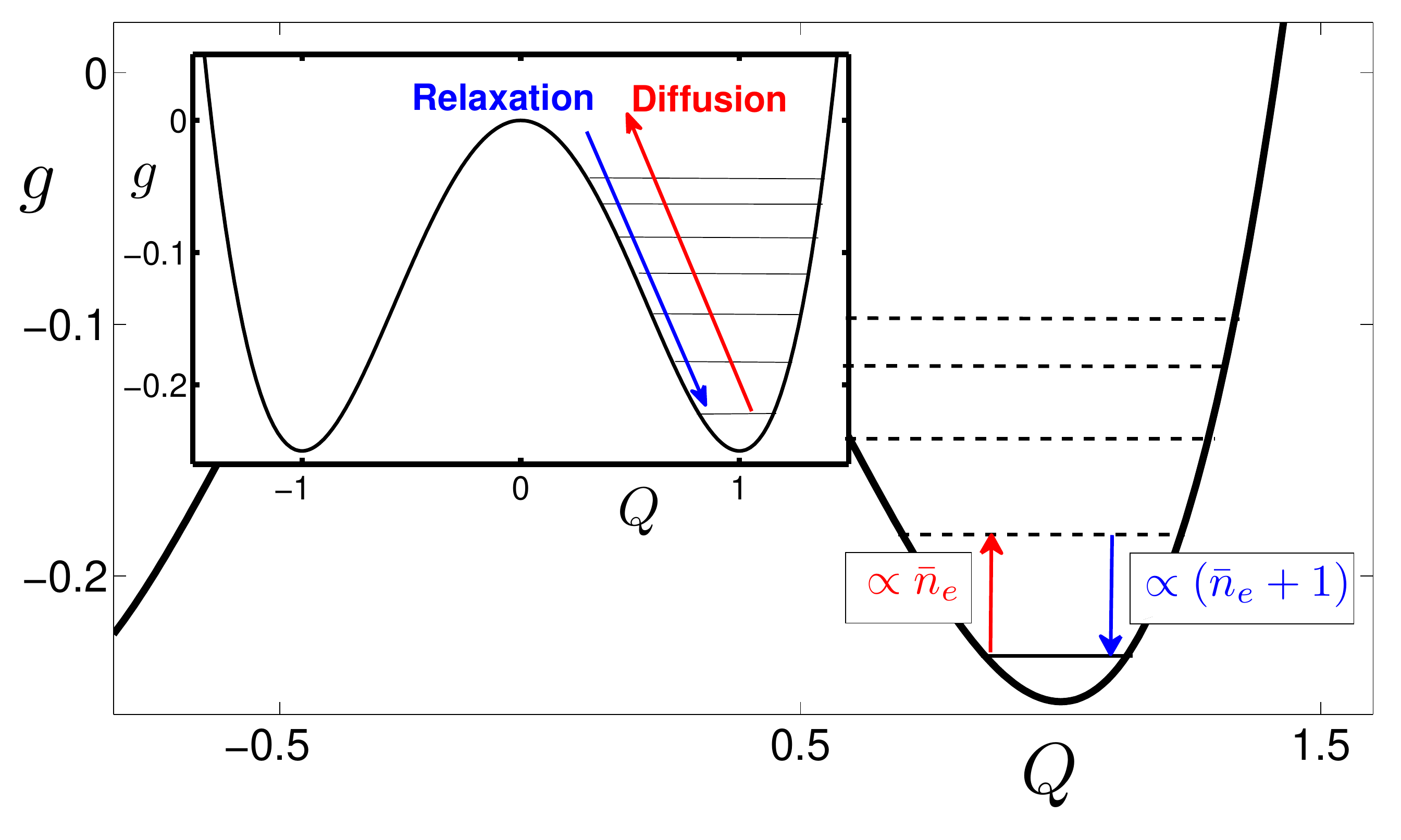}
\caption{The quasienergy Hamiltonian $g$ near its minimum as a function of the coordinate in the rotating frame $Q$ for $P=0$. The horizontal lines show schematically the quasienergy levels, which are weakly nonequidistant near the minimum of $g$. The vertical arrows show transitions that lead to the peaks in the power spectra with relative intensities that depend on the effective Planck number $\bar n_e$. The transitions between higher neighboring levels have close, but not identical frequencies, which can result in the onset of the fine structure of the noise spectrum. In the inset, the arrows indicate the change of $g$ in relaxation to and in diffusion away from the classically stable state. }
\label{fig:quasienergy}
\end{figure}

Quantum mechanically, a parametrically modulated oscillator can be naturally described in terms of the Floquet, or quasienergy states $\psi_{\ep}(t)$, such that $\psi_{\ep}(t+2\tau_F)=\exp(-2i\ep\tau_F/\hbar)\psi_{\ep}(t)$, where $\tau_F=2\pi/\omega_F$ is the modulation period and $\ep$ is the quasienergy. The quasienergy levels are sketched in Fig.~\ref{fig:quasienergy}. The lowest state in the figure corresponds to the stable state of parametrically excited vibrations, in the presence of weak coupling to the bath; there are two such states in the neglect of tunneling. If these were energy levels, for zero bath temperature $T$ the oscillator would be at the lowest level and its power spectrum would have one line that corresponds to the transition to the next level. However, because of quantum heating the oscillator occupies higher-$\ep$ states in Fig.~\ref{fig:quasienergy}. Therefore the spectrum has lines that correspond to transitions from these states as well, and the transitions not only up, but also down in $\ep$. Where different transitions are spectrally resolved, one gets direct information of the occupation of the states, and thus of the quantum temperature that characterizes the quasienergy distribution. We note that this temperature generally depends on quasienergy \cite{Dykman1988a,*Dykman2007,Marthaler2006}, but for a few lowest quasienergy states this dependence is not important.

\section{Oscillator dynamics in slow time}
\label{sec:slow_time}
\subsection{Quasienergy Hamiltonian in the rotating wave approximation}

The Hamiltonian of a parametrically modulated nonlinear oscillator is
\begin{equation}
\label{eq:H_0(t)}
H_0=\frac{1}{2}p^2+\frac{1}{2}q^2\left[\omega_0^2+F\cos(\omega_F
t)\right]+\frac{1}{4}\gamma q^4.
\end{equation}
Here, the mass is set equal to one, $F$ is the modulation amplitude, and
$\gamma$ is the anharmonicity parameter. We assume the modulation to be resonant and comparatively weak, $|F|\ll \omega_0^2$, so that the nonlinearity is also weak, which allows us to keep the lowest-order relevant nonlinear term in $H$,
\begin{eqnarray}
\label{eq:delta_omega}
|\omega_F-2\omega_0|\ll \omega_0\, ,\qquad |\gamma\langle q^2
\rangle|\ll\omega_0^2.\nonumber
 \end{eqnarray}
For concreteness we set $F,\gamma>0$.

Following \cite{Marthaler2006}, we change to the rotating frame using the canonical transformation $U(t)=\exp\left(-ia^{\dag}a\,\omega_Ft/2\right)$, where $a^{\dag}$ and $a$ are the raising and lowering operators of the oscillator, and introduce slowly varying in time dimensionless coordinate $Q$ and momentum $P$,
%
%\begin{eqnarray*}
%\label{eq:UdU}
%&& \\
$U^{\dag}(t)q U(t) = C_{\rm par}\left[P\cos(\omega_F t/2)-Q\sin(\omega_F t/2)\right]$,
%&&
$U^{\dag}(t)p U(t) = -(C_{\rm par}\omega_F/2)\left[P\sin(\omega_F t/2)+Q\cos(\omega_F
 t/2)\right]$,
%\end{eqnarray*}
%
where $C_{\rm par}=(2F/3\gamma)^{1/2}$. The commutation relation between $P$
and $Q$ has the form
\begin{equation}
\label{eq:L}
[P,Q]=-i\lambda\, ,\quad \lambda=3\gamma\hbar/F\omega_F\, .
\end{equation}
The dimensionless parameter $\lambda$ plays the role of the Planck constant in the quantum dynamics in the rotating frame. Respectively, the oscillator raising and lowering operators are expressed in terms of $P,Q$, and $\lambda$ in a standard way, %
\[U^{\dagger}(t)aU(t) =(2\lambda)^{-1/2}(P-iQ)\exp(-i\omega_Ft/2).\]

It is convenient to analyze the dynamics of a resonantly modulated weakly nonlinear oscillator in the rotating wave approximation (RWA). In this approximation the Hamiltonian in the rotating frame becomes $\tilde{H}_0\to U^{\dag}H_0 U-i\hbar U^{\dag}\dot{U}\approx
(F^2/6\gamma)\,\hat{g}$, where
\begin{eqnarray}
\label{eq:g}
\hat g &\equiv&  g(Q,P) =
\frac{1}{4}\left(P^2+Q^2\right)^2
+\frac{1}{2}(1-\mu)P^2\nonumber\\
&&-\frac{1}{2}(1+\mu)Q^2\, ,\qquad \mu= \frac{\omega_F(\omega_F-2\omega_0)}{F} .
%\nonumber
\end{eqnarray}
The dimensionless operator $\hat g$ describes the oscillator dynamics in slow dimensionless time, with the Schr\"odinger equation of the form of
\[ i\lambda\dot\psi \equiv i\lambda \partial_{\tau}\psi= \hat g \psi, \qquad \tau=Ft/2\omega_F.\]
The eigenvalues $g_n$ of $\hat g$ give the oscillator quasienergies $\ep_n= (F^2/6\gamma)g_n$.

Operator $\hat g$ does not have the form of a sum of kinetic and potential energies and depends on one dimensionless parameter $\mu$. We will consider region $-1< \mu < 1$ where function $g(Q,P)$ has two minima and a maximum. Its cross-section by the plane $P=0$ is shown in the inset of Fig.~\ref{fig:quasienergy}.

The full oscillator dynamics, including dissipation, can be described by the master equation for the oscillator density matrix $\rho$. For dissipation that comes from weak coupling to a thermal reservoir, which is linear in the oscillator coordinate and possibly momentum, for almost resonant modulation, $|2\omega_0-\omega_F|\ll \omega_F$, comparatively weak nonlinearity, $|\gamma|\langle q^2\rangle \ll \omega_0^2$, and the density of states of the reservoir weighted with the interaction smooth around $\omega_0$ this equation in the RWA has the form
\begin{eqnarray}
\label{eq:QKE_operator_form}
\dot\rho &&=i\lambda^{-1}[\rho,\hat g]-\hat\kappa\rho,\qquad
\hat\kappa\rho =\kappa (\bar{n}+1)(a^{\dagger}a\rho\nonumber\\
&&-2 a\rho a^{\dagger}+\rho a^{\dagger}a)
+\kappa \bar n (aa^{\dagger}\rho - 2 a^{\dagger}\rho a +\rho a a^{\dagger}).
\end{eqnarray}
Here, operator $\hat \kappa\rho$ describes dissipation. The transition to the interaction representation with respect to the oscillator variables is done using operators $U(t), U^{\dagger}(t)$. The renormalization of the oscillator frequency due to the bath is incorporated into $\omega_0$. The dimensionless parameter $\kappa =  2\omega_F\Gamma/F$ is proportional to the oscillator decay rate $\Gamma$; this rate gives the ring-down time $1/2\Gamma$ and the quality factor $\omega_0/2\Gamma$, which we assume to be large. We note that in Ref.~\onlinecite{Marthaler2006} we used $\eta$ instead of $\kappa$.

In the limit of small $\kappa$ the minima of $g(Q,P)$ correspond to the stable stationary states in the rotating frame, and thus to the stable states of period-two vibrations at frequency $\omega_F/2$, in the laboratory frame.

\subsection{Noise power spectrum}

Of significant interest for the experiment are spectra of modulated oscillator \cite{Dykman1979a,Drummond1980c}, including the power spectrum and the spectra of absorption of an additional weak field and radiation emission. Measurements of the power spectrum have been already reported \cite{Wilson2010} for a microwave cavity with length effectively modulated by a superconducting interference device \cite{Wallquist2006}; the related spectrum can be studied also through sideband absorption of a Josephson junction based qubit coupled to a driven nonlinear resonator [13]. The power spectrum of the oscillator also determines relaxation of a qubit coupled to it \cite{Serban2010}.

We will consider the power spectrum at frequencies close to the eigenfrequency of the parametrically modulated oscillator. In the vicinity of the maximum, this spectrum is given by
\begin{eqnarray}
\label{eq:spectrum_defined}
\Phi(\omega)={\rm Re}\int\nolimits_0^{\infty}dt e^{i\omega t}\langle\langle a(t)\ad(0)\rangle\rangle .
\end{eqnarray}
Here,
\[\langle\langle A(t)B(0)\rangle\rangle=\frac{\omega_F}{4\pi}\int\nolimits_0^{4\pi/\omega_F} dt_i\langle A(t+t_i)B(t_i)\rangle,\]
where $\langle\ldots\rangle$ indicates ensemble averaging.

For small fluctuation intensity, Eq.~(\ref{eq:lambda_limit}), the spectrum $Q(\omega)$ has distinct peaks near $\omega_F/2$. One or two of them, possibly with fine structure, see below, come from small-amplitude fluctuations about the stable states. In addition, there is an extremely narrow spectral peak from rare transitions between the states centered at frequency $\omega_F/2$. This peak is discussed in Sec.~\ref{sec:supernarrow}.

Here and in Secs.~\ref{sec:quantum_T}-\ref{sec:moderate_damping} we will be interested in the peak(s) of $\Phi(\omega)$ due to small-amplitude quantum and classical fluctuations about the classically-stable states of period-two vibrations. The coordinates of  these state $\pm (Q_0,P_0)$ in the rotating frame for arbitrary dimensionless decay rate $\kappa$ are determined in Sec.~\ref{sec:moderate_damping}, see Eq.~(\ref{eq:Q_0-P_0}); in the limit of small $\kappa$ the states are located at the minima of $g(Q,P)$, with $Q_0\approx (1+\mu)^{1/2}, P_0\approx 0$ from Eq.~(\ref{eq:g}). The states are symmetrical, since they correspond to time translation by the modulation period, in the laboratory frame. The contributions to $\Phi(\omega)$ from fluctuations about them are equal, and it is sufficient to study one of them. In doing so we will assume that the oscillator is localized in the vicinity of the stable state $(Q_0,P_0)$ and disregard interstate transitions. The corresponding term in $\Phi(\omega)$ is $\Phi_0(\omega)$, with
\begin{equation}
\label{eq:Phi_0_defined}
\Phi_{0}(\omega)= {\rm Re}\int\nolimits_0^{\infty}dt e^{i\omega t}\langle\langle \delta a(t)\delta a^{\dagger}(0)\rangle\rangle.
\end{equation}
Here, $\delta a(t) = a(t)- a_0(t)$ is the operator $a$ counted off from its expectation value $a_0$ at the stable state $(Q_0,P_0)$, \[a_0(t)=(2\lambda)^{-1/2}(P_0-iQ_0)\exp(-i\omega_Ft/2).\]

\section{Quantum temperature in the small damping limit}
\label{sec:quantum_T}
\subsection{The Bogoliubov transformation and the quasienergy spectrum}

Quantum noise is most clearly manifested in the spectrum if the oscillator relaxation rate is small, so that the relaxation-induced width of the quasi-energy levels exceeds the inter-level distance. For small $\lambda$ the dimensionless interlevel distance is $\lambda \nu(g)$, where $\nu(g)$ is the dimensionless frequency of classical vibrations described by equations $\dot Q=\partial_P g,\; \dot P=-\partial_Q g$ (in Ref.~\onlinecite{Marthaler2006} we used $\omega(g)$ instead of $\nu(g)$). The dimensionless level width is proportional to the decay rate $\kappa $. Therefore the condition of well-separated levels is $\nu(g)\ll \kappa $.

We are interested in the levels close to the minima of function $g(Q,P)$, see Fig.~\ref{fig:quasienergy}, i.e., for $g\approx g_{\min}=-(1+\mu)^2/4$. Then, from Eq.~(\ref{eq:g}) the condition of narrow levels has the form
\begin{equation}
\label{eq:weak_dissipation_classical}
\nu_0\gg \kappa , \qquad \nu_0\equiv \nu(g_{\min})= 2(1+\mu)^{1/2}.
\end{equation}
We assume that, at the same time, the level width largely exceeds the splitting due to resonant tunneling between the minima of $g$; this splitting is exponentially small for $\lambda\ll 1$.

Where these conditions are held, the oscillator motion near $g_{\min}$ is weakly damped vibrations at dimensionless frequency $\approx\nu_0$. It can be studied using the Bogoliubov transformation from $a,\ad$ to new operators $b,b^{\dagger}$,
\begin{eqnarray}
\label{eq:Bogoliubov_transform}
&&  U^{\dagger}(t)aU(t)=a_0(t)+(u b+v b^{\dag})e^{-i\omega_Ft/2}, \nonumber\\
&&  u= -i(2\nu_0)^{-1/2}\left(1+\frac{\nu_0}{2}\right), \nonumber\\
&&v=-i(2\nu_0)^{-1/2}\left(1-\frac{\nu_0}{2}\right);
 \end{eqnarray}
The coefficients $u,v$ are chosen so that, to second order in $P, Q-Q_0$, $\hat g \approx \lambda \nu_0 b^{\dagger}b + $~const, that is, near its minimum $\hat g$ becomes the Hamiltonian of an auxiliary harmonic oscillator with dimensionless frequency $\nu_0$. Operators $b$ and $b^{\dagger}$ are, respectively, the lowering and raising operators for this auxiliary oscillator. We emphasize that vibrations of this oscillator occur in the rotating frame and correspond to the vibrations of the original oscillator at frequencies $\omega_F/2 \pm (F/2\omega_F)\nu_0$. We note that the Bogoliubov transformation can be written as a squeezing transformation, 
\begin{equation}
\label{eq:squeezing}
(2\lambda)^{-1/2}(Q-Q_0+iP)= b\cosh r_* - b^{\dagger}\sinh r_*
\end{equation}
with $\cosh r_*=iu$ and $\sinh r_* = -iv$. 

Higher-order terms in $P, Q-Q_0$ in $\hat g$ lead to anharmonicity of vibrations about $g_{\min}$. In turn, the anharmonicity leads to nonequidistance of the vibrational energy levels, that is, of the quasienergy levels of the original oscillator. To the lowest order, the nonequidistance is determined by the terms quadratic in $b^{\dagger}b$ taken to the first order and by the cubic terms in $b,b^{\dagger}$ taken to the second order. This gives for the eigenvalues of $\hat g$
\begin{equation}
\label{eq:g_eigenvalues}
g_n\approx \lambda \tilde\nu_0 n+\frac{1}{2}\lambda^2 V n^2+\tilde g_{\rm min}, \qquad
V=- \frac{\mu + 4}{\mu+1}.
\end{equation}
Here, $\tilde\nu_0=\nu_0+\lambda V/2$ and $\tilde g_{\min} -g_{\min}\sim \lambda$. Parameter $V$ gives the nonequidistance of the levels, with the transition frequencies forming a ladder, $\nu(g_n)=(g_{n+1}-g_n)/\lambda = \nu_0+\lambda V(n+1)$. We note that the frequency step $\lambda V$ is proportional to the anharmonicity parameter $\gamma$ of the original oscillator. Equation (\ref{eq:g_eigenvalues}) applies for small $\lambda$ and small $n$, where $\lambda |V|n\ll \nu_0$.

\subsection{Master equation in terms of the transformed operators}

The full oscillator dynamics near the minima of $g(Q,P)$ can be described by the master equation (\ref{eq:Liouville_compact}) with $a^{\dagger},a$ written in terms of the operators $b^{\dagger}, b$. For small $\kappa$ the master equation can be simplified by noting that, for $\kappa=0$, matrix elements of $\rho$ on the eigenfunctions $|n\rangle$ of $\hat g$ oscillate in dimensionless time as $\rho_{mn}\propto \exp[-i\nu_0(m-n)\tau/\lambda]$, for small $n,m$. Dissipation couples matrix elements $\rho_{mn}$ with $\rho_{m'n'}$. For weak dissipation, where Eq.~(\ref{eq:weak_dissipation_classical}) holds, the coupling is particularly strong if $m-n = m'-n'$. Such coupling is described if in $\hat\kappa\rho$ written in terms of $b^{\dagger},b$ we keep terms with equal numbers of $b$ and $b^{\dagger}$ operators, whereas the terms that contain $b^2$ and $(b^{\dagger})^2$ are disregarded. The resulting expression has the form
\begin{eqnarray}
\label{eq:dissipation_with_b_b+}
\hat\kappa\rho&&=
           \kappa (\bar{n}_e+1)(b^{\dag}b\rho - 2 b\rho b^{\dag} +\rho b^{\dag}b)\nonumber\\
         & &+\kappa \bar{n}_e(b b^{\dag}\rho - 2 b^{\dag}\rho b + \rho b b^{\dag}).
\end{eqnarray}
with
\begin{eqnarray}
\label{eq:n_eff}
\bar{n}_e &=& \bar n + (1+2\bar n)\sinh^2r_*\nonumber\\
&&=\left[(\mu+2)(2\bar{n}+1)-\nu_0\right]/2\nu_0.
\end{eqnarray}
From the comparison of Eqs.~(\ref{eq:QKE_operator_form}) (\ref{eq:dissipation_with_b_b+}) one can see that $\bar{n}_e$ plays the role of the effective Planck number for vibrations about $g_{\min}$. 

The stationary solution of the master equation near the chosen minimum of $g$ has the form of the Boltzmann distribution, $\rho^{\rm (st)}_{mm}\propto [\bar{n}_e/(\bar{n}_e+1)]^m$ or, in the operator form,
\begin{eqnarray}
\label{eq:rho_st}
&&\rho^{\rm (st)}=(\bar n_e+1)^{-1}\exp(-\lambda\nu_0b^{\dagger}b/{\cal T}_e),\\
&&{\cal T}_e=\lambda\nu_0 /\ln[(\bar{n}_e+1)/\bar{n}_e]. \nonumber
\end{eqnarray}
Here, ${\cal T}_e$ is the dimensionless effective temperature of vibrations about $g_{\min}$. Equation (\ref{eq:n_eff}) coincides with the results \cite{Marthaler2006} obtained by a completely different method. For $\bar n=0$ the result coincides also with what follows from the analysis of a different model of a modulated oscillator \cite{Peano2010,*Peano2010a} if one uses the appropriate value $r_*$ of the squeezing transformation (\ref{eq:squeezing}). The normalization of $\rho^{\rm (st)}$ corresponds to the assumption that the oscillator is localized in the vicinity of the stable state $(Q_0,P_0)$. The distribution over quasienergy states for other systems and other relaxation mechanisms were discussed recently in Refs.~\onlinecite{Verso2010,Ketzmerick2010}.

\begin{figure}[h]
\begin{center}
\includegraphics[width=8.5 cm]{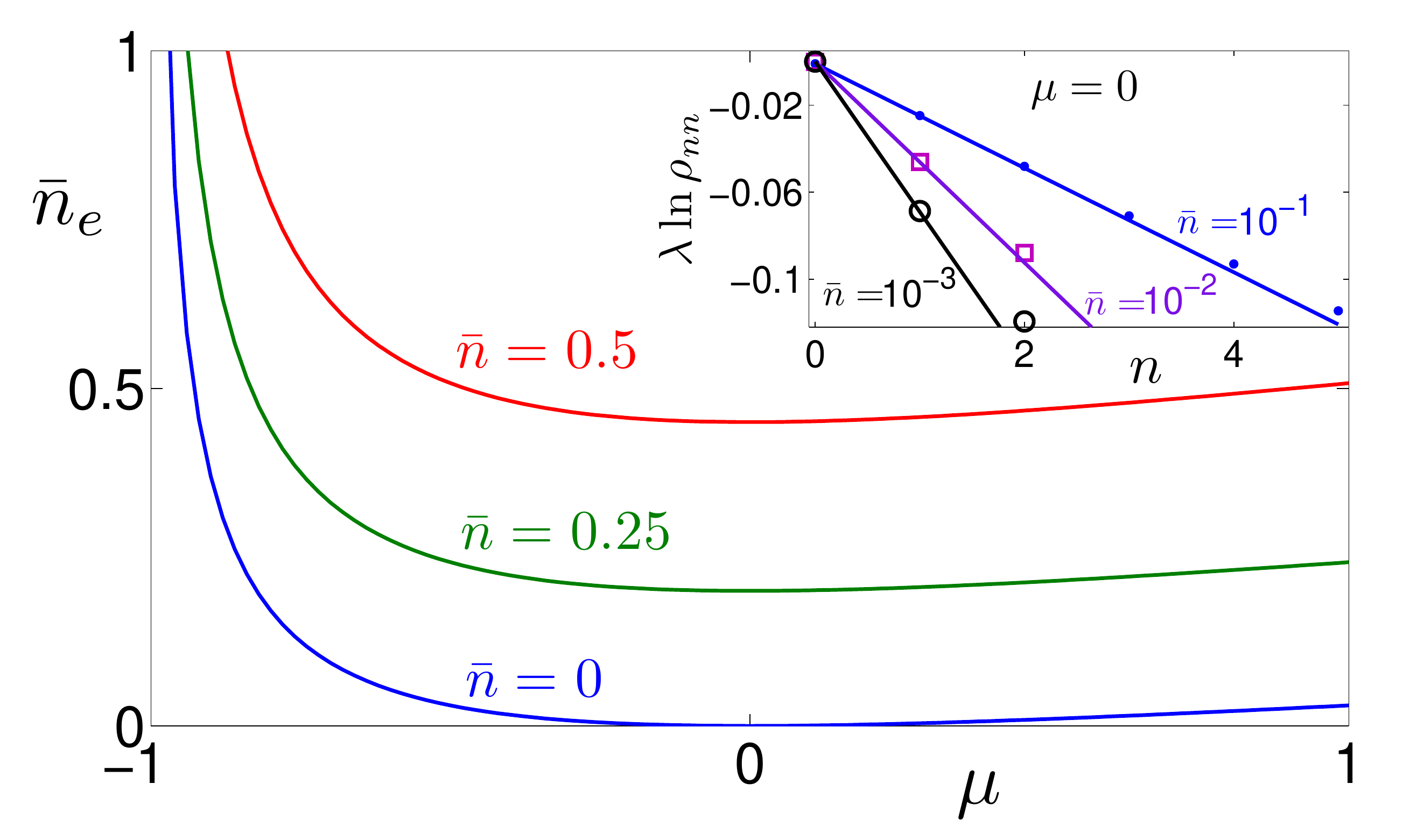}
\caption{The effective oscillator Planck number $\bar n_e$ as a function of the scaled frequency detuning $\mu=\omega_F(\omega_F-2\omega_0)/F$ for different values of the Planck number $\bar n$, for small decay rate. The inset shows the populations of the low-lying quasienergy states $\rho_{nn}$ for $\mu=0$. The solid lines in the inset show the results of the calculation that takes into account relaxation-induced transitions between neighboring quasienergy levels, Eq.~(\ref{eq:n_eff}), shown in the main figure. The dots, squares, and circles show the result of the full calculation, Ref.~\onlinecite{Marthaler2006}.}
\label{fig:occupation}
\end{center}
\end{figure}

It follows from Eq.~(\ref{eq:n_eff}) that the effective Planck number, and thus also ${\cal T}_e$, remain nonzero even for zero temperature of the bath, $\bar n=T=0$. This is a consequence of quantum fluctuations that accompany oscillator relaxation, and therefore we call ${\cal T}_e$ quantum temperature. The dependence of $\bar n_e$ on the dynamical parameter $\mu$ is shown in Fig.~\ref{fig:occupation}. It is nonmonotonic, with a minimum at exact resonance between the driving frequency and twice the oscillator eigenfrequency, where $\mu\propto \omega_F - 2\omega_0=0$. The value of $\bar n_e$ increases with decreasing $\mu+1\propto \nu_0^2$, i.e., close to the bifurcation point where the period-two vibrations are excited. However, the assumption $\nu_0\gg \kappa$ breaks down sufficiently close to the bifurcation point, which imposes a restriction on $\bar n_e$.

The occurrence of the minimum of $\bar n_e$ is an interesting feature of the parametrically modulated oscillator. For $\mu=0$, from Eq.~(\ref{eq:n_eff})  $\bar{n}_e=\bar n$; the effective temperature is equal to the thermal bath temperature, and ${\cal T}_e=0$ for $T=0$. We note that this is an asymptotic result that applies only very close to $g_{\min}$. The stationary distribution $\rho_{nn}^{\rm (st)}$ is Boltzmann-like only to the leading order in the distance $g_n-g_{\min}\ll g_{\min}$. Strictly speaking, the effective temperature is quasienergy-dependent \cite{Marthaler2006}. Even for $\mu=T=0$ quasienergy states with $n\geq 1$ are occupied, but in the range of small $n$ this occupation is much smaller for $\mu=0$ than for $|\mu|\sim 1$. The role of the corrections to the Boltzmann distribution for $\mu=0$ is illustrated in the inset of Fig.~\ref{fig:occupation}.

\section{Fine structure of the power spectrum}
\label{sec:fine_structure}
\subsection{General expression for the spectrum near its maximum}

Equations (\ref{eq:g_eigenvalues}) - (\ref{eq:dissipation_with_b_b+}) describe the dynamics of the modulated underdamped oscillator in terms of an auxiliary oscillator in thermal equilibrium with temperature ${\cal T}_e$. The power spectrum $\Phi_0(\omega)$ of the original oscillator near its maxima can be expressed in terms of the power spectrum of the auxiliary oscillator. From Eqs.~(\ref{eq:Phi_0_defined}) and (\ref{eq:Bogoliubov_transform}), for frequencies $\omega-\omega_F/2$ close to the frequency of vibrations about the stable states,
\begin{eqnarray}
\label{eq:spectra_via_b_b+}
&&\Phi_0(\omega)\approx \frac{2\omega_F}{F}|u|^2\Phi_b(\nu), \quad |\nu-\nu_0|\ll \nu_0\\
&&\Phi_b(\nu)={\rm Re}~\int\nolimits_0^{\infty}d\tau e^{i\nu\tau}\langle\langle b(\tau)b^{\dagger}(0)\rangle\rangle_{\rm rot}.\nonumber
\end{eqnarray}
Here, $\nu$ is the dimensionless frequency counted off from $\omega_F/2$,
\[ \nu=(2\omega_F/F)\left(\omega- \omega_F/2\right);\]
the subscript in $\langle\langle \ldots\rangle\rangle_{\rm rot}$ indicates that the correlator is calculated in the rotating frame,
\begin{equation}
\label{eq:correlator_defined}
\langle\langle A(\tau)B(0)\rangle\rangle_{\rm rot} = {\rm Tr}~A\rho(\tau;B), \qquad \rho(0;B)=B\rho^{\rm (st)},
\end{equation}
where $\rho(\tau;B)$ satisfies the master equation (\ref{eq:QKE_operator_form}) with the dissipative term of the form Eq.~(\ref{eq:dissipation_with_b_b+}) and $\rho^{\rm (st)}$ is the stationary distribution given by Eq.~(\ref{eq:rho_st}).

In deriving Eq.~(\ref{eq:spectra_via_b_b+}) we took into account that, if we ignore dissipation and the nonlinearity of vibrations about $g_{\min}$, $\exp(ig\tau/\lambda)b\exp(-ig\tau/\lambda)\approx \exp(-i\nu_0\tau)b$, and therefore function $\Phi_b(\nu)$ describes the dominating contribution to $\Phi_0$ for $\nu$ close to $\nu_0$. Using the fact that small-amplitude vibrations near $g_{\min}$ can be thought of as being close to equilibrium, one can show that the peak of $\Phi_0$ for $\nu$ close to $-\nu_0$ is described by function
$(2\omega_F|v|^2/F)\exp(-\lambda\nu_0/{\cal T}_e)\Phi_b(-\nu)$.

\subsection{Effective partial spectra representation}

The problem of the power spectrum of a weakly nonlinear underdamped oscillator was discussed previously \cite{Dykman1973}. Applying the results to the spectrum $\Phi_b(\nu)$ of the auxiliary oscillator, after some straightforward transformations we obtain
\begin{eqnarray}
\label{eq:partial_general}
 &&2\Phi_b(\nu) = (\bar n_e+1){\rm Re}~\sum_n\phi_b(n,\nu);\\
 &&\phi_b(n,\nu)=
 4n(\Lambda-1)^{n-1}(\Lambda+1)^{-(n+1)}\nonumber\\
 &&\times \left[\kappa (2 \aleph n -1 -i\vartheta)-i(\nu-\tilde\nu_0)\right]^{-1}\nonumber
\end{eqnarray}
where
\begin{eqnarray}
\label{eq:parameters_partial_spectr}
\Lambda& = &\aleph^{-1}\left[1 +i\vartheta(2\bar{n}_e+1)\right]\, , \qquad \vartheta =\lambda V/2\kappa,\nonumber\\
\aleph  &=& \left[1 + 2i \vartheta(2\bar{n}_e+1)-\vartheta^2\right]^{1/2}\, \quad ({\rm Re}~\aleph > 0).\nonumber
\end{eqnarray}

Equation (\ref{eq:partial_general}) can be thought of as a representation of the spectrum as a sum of  effective partial spectra Re~$\phi_b(n,\nu)$ that correspond to transitions $n-1\to n$ between the oscillator quasienergy levels. Functions $\phi_n(n,\nu)$ depend on two parameters, $\vartheta$ and $\bar n_e$. Parameter $\vartheta$ gives the ratio of the difference $\lambda V= \nu(g_{n})-\nu(g_{n-1})$ between neighboring transition frequencies and the broadening $\kappa $ of the quasienergy levels, whereas the effective Planck number $\bar n_e$ gives the typical width of the stationary distribution over the levels.

The form of $\phi_b(n,\nu)$ is particularly simple for a comparatively large frequency spacing or small damping, $\lambda |V|\gg \kappa $. Note that for small $\lambda$ this is a much stronger restriction on the decay rate than the condition $\kappa \ll \nu_0$ used to derive Eq.~(\ref{eq:partial_general}). For such a small decay rate
\begin{eqnarray}
\label{eq:fine_structure}
\phi_b(n,\nu)&\approx & \frac{n}{\bar n_e+1}e^{-\lambda\nu_0(n-1)/{\cal T}_e}\nonumber\\
&&\left\{\kappa_n  -i\left[\nu-\nu(g_{n-1})\right]\right\}^{-1},\quad |\vartheta|\gg 1,\nonumber\\
&&\kappa_n   = \kappa [2n(2\bar n_e+1)-1].
\end{eqnarray}
In this limit Re~$\phi_b(n,\nu)$ is a Lorentzian line centered at the frequency $\nu(g_{n-1})=\nu_0+\lambda V n$ of transition $n-1 \to n$,  with halfwidth $\kappa_n  $ equal to the half-sum of the reciprocal lifetimes of the levels $n-1$ and $n$. Associating $\phi_b(n,\nu)$ with a partial spectrum is fully justified in this limit. The amplitude of $\phi_b(n,\nu)$ contains the Boltzmann factor proportional to the population of the quasienergy level $n-1$. The overall spectrum $\Phi_b(\nu)$ has a pronounced fine structure for $|\vartheta|\gg 1$. The line intensities immediately give the effective quantum temperature ${\cal T}_e$.

As $|\vartheta|$ decreases the partial spectra start to overlap and for $|\vartheta|\lesssim 1$ they can no longer be identified. Indeed, the typical dimensionless time $\lesssim\kappa^{-1}$ the oscillator spends in a given quasienergy state becomes smaller than the distance $|\lambda V|^{-1}$ between different transition frequencies $\nu(g_n)$. Therefore such frequencies cannot be resolved. In the limit $|\vartheta|\to 0$ we have $\phi_n(n,\nu)\propto \delta_{n,1}$, and the spectrum has the form of a single Lorentzian peak of dimensionless width $\kappa$,
\begin{equation}
\label{eq:Lorentzian}
\Phi_b(\nu)=(\bar n_e+1)~{\rm Re}~\left[\kappa - i(\nu-\nu_0)\right]^{-1}.
\end{equation}

\begin{figure}[h]
\includegraphics[width=8.5truecm]{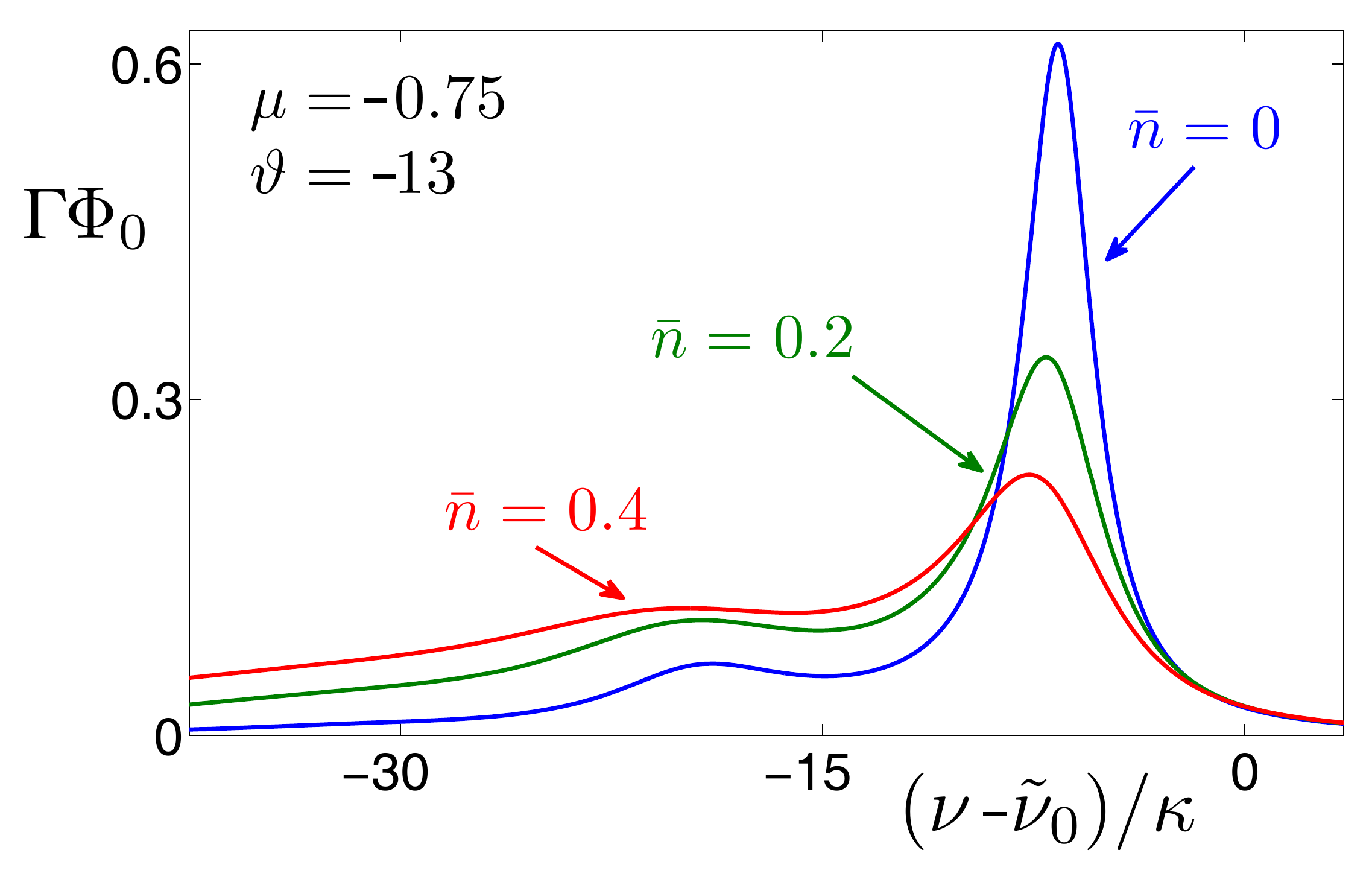}
\caption{The fine structure of the noise spectrum of a modulated oscillator near resonant frequency $\omega_F/2 + F\tilde\nu_0/2\omega_F$ [$\nu/\kappa=(\omega-
\omega_F/2)/\Gamma$]. The intensities of the fine structure lines are determined by the effective Planck number $\bar n_e$. Due to quantum heating, the fine structure is seen even for zero bath temperature. The widths of the individual lines increase with increasing oscillator Planck number $\bar n$, leading to the smearing of the fine structure.}
\label{fig:fine_structure}
\end{figure}

The evolution of the spectrum with varying $\bar n_e$ depends on $|\vartheta|$. For $|\vartheta|\ll 1$ the spectrum is Lorentzian for $\bar n_e \lesssim 1$, but becomes non-Lorentzian and asymmetric for large $\bar n_e$, where $|\vartheta|\bar n_e > 1$. However, the spectrum does not have a fine structure in this case. On the other hand, for $|\vartheta|\gg 1$ a fine structure emerges, but only in a limited range of $\bar n_e$. It is seen from Eq.~(\ref{eq:fine_structure}) that for $\bar n_e\ll 1$ only $\phi_b(1,\nu)$ has an appreciable intensity, $\phi_b(n\neq 1,\nu)$ are small. On the other hand, for large $\bar n_e$ the linewidth $\kappa_n$ becomes large and the spectral lines with different $n$ overlap, starting with large $n$. The evolution of the fine structure with varying $\bar n$ is illustrated in Fig~\ref{fig:fine_structure}. The quantitative results confirm the above qualitative arguments.

\section{Moderate damping}
\label{sec:moderate_damping}

\subsection{Master equation in the Wigner representation}

The oscillator power spectrum should be analyzed differently if the oscillator damping is not that small. We assume that the original oscillator remains underdamped, $F\kappa/\omega_F \ll \omega_0$, but for the auxiliary oscillator which vibrates about the minimum of $g(Q,P)$, the dimensionless decay rate $\kappa$ exceeds the nonequidistance of the quasienergy levels, $\kappa\gg \lambda|V|$, so that the power spectrum does not have the fine structure discussed in Sec.~\ref{sec:quantum_T}. The latter condition indicates that the quantum effects related to the difference of the transitions frequencies are small, since $\lambda \propto \hbar$. However, other quantum effects are still important, as seen below, and in particular the spectrum strongly depends on the quantum temperature. We note that, for $\kappa\gg \lambda|V|$, the ratio of the width of the quasienergy levels to the distance between them $\propto \kappa/\nu_0$ can still be arbitrary.

The analysis of the power spectrum for moderate damping can be done by writing master equation (\ref{eq:QKE_operator_form}) in the Wigner representation,
\begin{eqnarray}
\label{eq:Liouville_compact} \dot\rho_W=-\n\left({\bf
K}\rho_W\right) +\lambda \hat L^{(1)}\rho_W +\lambda^2\hat
L^{(2)}\rho_W.
\end{eqnarray}
Here, $\rho_W$ is the density matrix in the Wigner representation,
\begin{eqnarray*}
\label{eq:Wigner_def}
\rho_W( Q,P;\tau)=\int d\xi e^{-i\xi
P/\lambda}\rho\left( Q+\frac{1}{2}\xi, Q-\frac{1}{2}\xi;\tau\right),
\end{eqnarray*}
where $\rho(Q_1,Q_2;\tau)=\langle Q_1|\rho(\tau)| Q_2\rangle$ is the density matrix in the coordinate representation. In Eq~(\ref{eq:Liouville_compact}) we use vector notations, ${\bf K} = (K_Q,K_P)$ and $\n = (\partial_Q,\partial_P)$.

Vector ${\bf K}$ determines the evolution of the density matrix in
the absence of quantum and classical fluctuations,
\begin{eqnarray}
\label{eq:K_vector} K_Q=\partial_P g-\kappa Q \qquad
K_P=-\partial_Q g-\kappa P,
\end{eqnarray}
whereas the terms $\propto \lambda$ in Eq.~(\ref{eq:Liouville_compact}) account for fluctuations. If we set $\lambda = 0$, Eq.~(\ref{eq:Liouville_compact}) will describe classical motion
%
%\begin{equation}
%\label{eq:classical}
$\dot Q=K_Q,\; \dot P=K_P.$
%\end{equation}
%
The condition ${\bf K}={\bf 0}$ gives the values of $Q, P$ at the
stationary states of the oscillator in the rotating frame. For $|\mu| < (1-\kappa^{2})^{1/2}$ the system has 3 stationary states. One is located at $Q=P=0$ and is unstable. The other two are located symmetrically at $\pm (Q_0,P_0)$ with $Q_0=r_0\cos\theta$, $P_0=r_0\sin\theta$, where
\begin{equation}
\label{eq:Q_0-P_0}
r_0^2\equiv  Q_0^2+P_0^2=\mu + (1-\kappa^{2})^{1/2}
\end{equation}
and $\theta=\arctan\left\{\left[1-(1-\kappa^2)^{1/2}\right]/\kappa\right\}$. These states are asymptotically stable. Respectively, the real parts of the eigenvalues of matrix $\hat{\cal K}$,
\begin{equation}
\label{eq:K_matrix}
{\cal K}_{ij}=\left[\partial K_i/\partial X_j\right]_{Q_0,P_0} \qquad (X_1\equiv Q,\;X_2\equiv P)
\end{equation}
are negative [the subscript $(Q_0,P_0)$ indicates that $\hat{\cal K}$ is calculated at point $(Q_0,P_0)$]. For $\kappa \ll 1$ the stable states correspond to the minima of $g(Q,P)$ in Fig.~\ref{fig:quasienergy}.

The terms $\hat L^{(1)}$ and $\hat L^{(2)}$ in Eq.~(\ref{eq:Liouville_compact})
describe, respectively, the fluctuations due to decay processes and purely quantum fluctuations that are not related to the coupling to a thermal bath,
\begin{eqnarray}
\label{eq:L1L2}
&&\hat L^{(1)} = \kappa \left(\bar
n+1/2\right)\n^2 , \nonumber\\
&&\hat L^{(2)} =
-\frac{1}{4}\left(Q\partial_P-P\partial_Q\right)\n^2.
\end{eqnarray}
The decay-related fluctuations lead to diffusion in $(Q,P)$-space, as seen from
the structure of $\hat L^{(1)}$. In contrast, the term $\hat L^{(2)}$ is independent of $\kappa $ and contains third derivatives; for small $\lambda$ it is not important close to the stable states.

It follows from Eqs.~(\ref{eq:Liouville_compact}) and (\ref{eq:L1L2}) that, in the Wigner representation, the stationary distribution $\rho_W^{\rm (st)}$ has Gaussian peaks at the stable states $\pm (Q_0,P_0)$. They are of the same form for the both states, and close to $(Q_0,P_0)$
\begin{eqnarray}
\label{eq:Gaussian}
&&\rho_W^{\rm (st)}(Q,P)=\frac{(\det \hat A)^{1/2}}{2\bar n + 1}\exp\left[-\frac{\delta\Xb\hat A \delta\Xb}{\lambda(2\bar n +1)}\right],\nonumber\\
&&2\kappa \hat A^2 + \hat A\hat{\cal K} + \hat{\cal K}^{\dag}\hat A=0,
\end{eqnarray}
where
\[ \delta\Xb= (\delta Q,\delta P)\equiv (Q-Q_0,P-P_0)\]
is the distance from the stable state, $|\delta \Xb|^2\ll Q_0^2+P_0^2$. Equation (\ref{eq:Gaussian}) shows that the condition for quantum and classical fluctuations to be small is
\begin{equation}
\label{eq:lambda_limit}
\lambda (2\bar n + 1)\ll r_0^2.
\end{equation}
From Eqs.~(\ref{eq:K_vector}), (\ref{eq:K_matrix}), and (\ref{eq:Gaussian}) Tr~$\hat A = 2$ is independent of the parameters of the system. However, the distribution (\ref{eq:Gaussian}) is squeezed \cite{Walls2008}, and the variances of $Q-Q_0$ and $P-P_0$ depend on the oscillator parameters.

It is useful to note that, in the small-damping limit $\kappa\ll \nu_0$, matrix $\hat A$ is diagonal, with $A_{11}\approx 2(1+\mu)/(2+\mu)$, $A_{22}\approx 2/(2+\mu)$, and $A_{12}\propto\kappa/\nu_0$. Using the relation $2\bar n_e+1 =(\mu + 2)(2\bar n + 1)/\nu_0$ that follows from Eq.~(\ref{eq:n_eff}), one can see that the above expression for $\hat A$ leads to $\rho_W^{\rm (st)}\propto \exp\{-2[g(Q,P)-g_{\min}]/[\lambda \nu_0(2\bar n_e+1)]\}$, which is the standard form of the Wigner distribution of a harmonic oscillator; in the present case, the result refers to the auxiliary oscillator discussed in Sec.~\ref{sec:quantum_T}, with Hamiltonian $g(Q,P)$, frequency $\nu_0$, and Planck number $\bar n_e$. The result is fully consistent with what was found in Sec.~\ref{sec:quantum_T} using a different method.

\subsection{Power spectrum for moderate damping}

Equations (\ref{eq:Liouville_compact}) and (\ref{eq:Gaussian}) allow one to find the oscillator power spectrum for an arbitrary relation between the width of the quasienergy levels and the level spacing $\kappa/\nu_0$ but for
$\lambda |V|\ll \kappa$. Again, we will be interested in the contribution to the spectrum from small-amplitude fluctuations about the stable states. The general expression for this contribution follows from Eqs.~(\ref{eq:Phi_0_defined}) and (\ref{eq:correlator_defined}),
\begin{eqnarray}
\label{eq:power_spectr_Wigner_general}
\frac{F}{2\omega_F}\Phi_0(\omega)&=&{\rm Re}\int_0^{\infty}d\tau e^{\i\nu\tau}\int \frac{dQ\,dP}{4\pi\lambda^2} (\delta P-i\delta Q)\nonumber\\
&&\times\rho_W^{(+)}(Q,P;\tau),
\end{eqnarray}
where function $\rho_W^{(+)}$ satisfies master equation (\ref{eq:Liouville_compact}) with the initial condition
\begin{eqnarray}
\label{eq:initial_cond_Wigner}
\rho_W^{(+)}(Q,P;0)&=&2\left[\delta P+i\delta Q - \frac{1}{2}\lambda(i\partial_Q + \partial_P)\right]\nonumber\\
&&\times \rho_W^{(\rm st)}(Q,P)
\end{eqnarray}
Function $\rho_W^{(+)}(Q,P;0)$ is the Wigner-transform of the operator $(\delta\hat P+i\delta\hat Q)\hat \rho(\tau=0)$; for operator $\hat \rho$ this transform is defined by Eq.~(\ref{eq:Wigner_def}). Factor 2 in Eq.~(\ref{eq:initial_cond_Wigner}) accounts for the contribution of fluctuations about the state $-(Q_0,P_0)$; we have also taken into account in Eq.~(\ref{eq:power_spectr_Wigner_general}) that $\int dQ\,dP\rho_W = 2\pi\lambda$.

The calculation of the power spectrum using Eqs.~(\ref{eq:power_spectr_Wigner_general}) and (\ref{eq:initial_cond_Wigner}) is similar to that performed in the classical \cite{Dykman1979a,*Dykman1994b} and quantum theory \cite{Serban2010} for the power spectrum of an oscillator modulated by an additive force at frequency close to $\omega_0$. One should replace $\rho_W$ in Eq.~(\ref{eq:Liouville_compact}) with $\rho_W^{(+)}$, set ${\bf K}\approx \hat{\cal K}\delta\Xb$, multiply the equation by $\exp(i\nu\tau)$ and then in turns by $\delta P$ and $\delta Q$. One should then integrate the resulting equation over $\tau, P, Q$, as in Eq.~(\ref{eq:power_spectr_Wigner_general}). This will lead to two coupled linear equations for the Fourier transforms of $\langle\delta P(\tau)[\delta P(0)+i\delta Q(0)]\rangle$ and $\langle\delta Q(\tau)[\delta P(0)+i\delta Q(0)]\rangle$. The inhomogeneous parts of these equations are determined by the average values $\langle\delta X_i\delta X_j\rangle$, which can be found from Eq.~(\ref{eq:Gaussian}). A straightforward but cumbersome calculation gives
\begin{widetext}
\begin{eqnarray}
\label{eq:spectrum_moderate_damping}
\frac{F}{2\omega_F}\Phi_0(\omega)
=\kappa
\frac{(\bar n +1)\left[\left(\nu +2r_0^2-\mu\right)^2 + \kappa^2\right] + \bar n(1+ r_0^4-\nu_a^2/2)}
{(\nu^2-\nu_a^2)^2+4\kappa^2\nu^2}, \qquad \nu=\frac{\omega_F(2\omega-\omega_F)}{F},
\end{eqnarray}
\end{widetext}
where $r_0$ is the dimensionless amplitude of parametrically excited vibrations in the neglect of fluctuations given by Eq.~(\ref{eq:Q_0-P_0}). The frequency $\nu_a=2r_0(r_0^2-\mu)^{1/2}$ characterizes damped vibrations about the stable state in the absence of fluctuations, $\nu_a^2=-\det \hat{\cal K} >0$.

For small but not too small damping, $\lambda |V|\ll \kappa \ll \nu_0$ we have $\nu_a\approx \nu_0$. One can then show from Eq.~(\ref{eq:spectrum_moderate_damping}) that function $\Phi_0(\omega)$ has two Lorentzian peaks at dimensionless frequencies $\pm\nu_0$ with halfwidth $\kappa$. The expression for the peak at $\nu_0$ coincides with Eqs.~(\ref{eq:spectra_via_b_b+}) and (\ref{eq:Lorentzian}), whereas for $\nu$ close to $-\nu_0$
\begin{equation}
\label{eq:anti_Stokes}
\frac{F}{2\omega_F}\Phi_0(\omega)\approx |v|^2\bar n_e \kappa \left[(\nu + \nu_0)^2+\kappa^2\right]^{-1},
\end{equation}
in agreement with Sec.~\ref{sec:fine_structure}. We emphasize that, in the laboratory frame, the spectral peaks described by Eqs.~(\ref{eq:Lorentzian}) and (\ref{eq:anti_Stokes}) lie on the opposite sides of frequency $\omega_F/2$ at the distance $F\nu_0/2\omega_F$ in dimensional frequency. The ratio of their intensities is proportional to the factor $\exp(-\lambda\nu_0/{\cal T}_e)$ and thus strongly depends on the quantum temperature, which provides an independent means for measuring this temperature. Even for zero temperature of the thermal reservoir ${\cal T}_e > 0$, generally, and therefore both peaks are present in the spectrum.

\begin{figure}
\includegraphics[width=8.5truecm]{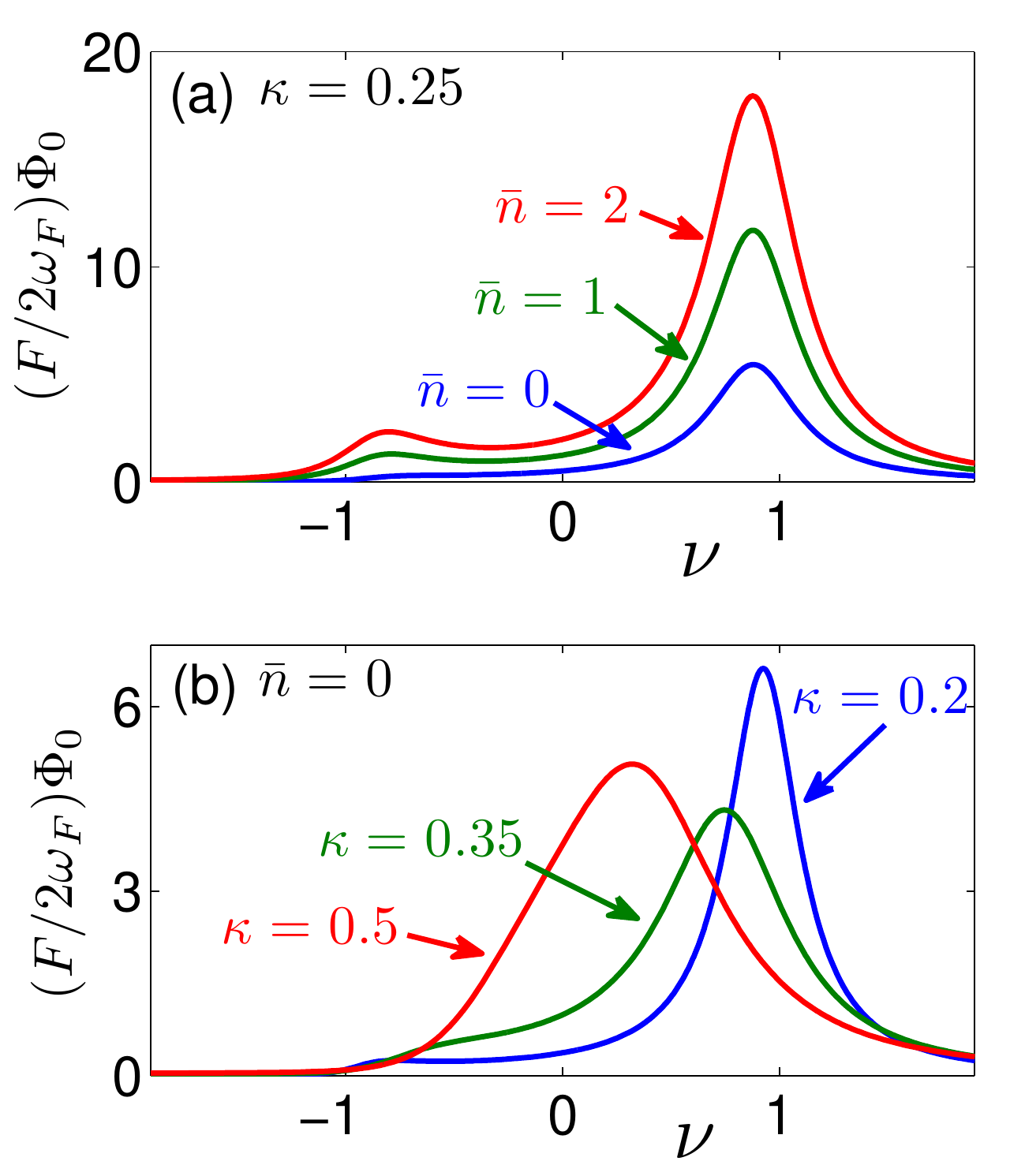}
\caption{The scaled contribution to the power spectrum from small-amplitude fluctuations about the stable period-two states $(F/2\omega_F)\Phi_0(\omega)$ for moderate damping where the fine structure is smeared out. The data refer to $\mu=-0.75$. Panels (a) and (b) show the evolution of the spectra with varying bath temperature, which determines the Planck number $\bar n$, and with the dimensionless decay rate $\kappa$, respectively. For comparatively small $\kappa$ the spectrum has two peaks, which in the laboratory frame are located at $\approx \omega_F/2 \pm F\nu_a/2\omega_F$. They correspond to transitions up and down in quasienergy, cf. Fig.~\ref{fig:quasienergy}. The peak at $\omega_F/2 - F\nu_a/2\omega_F$ for zero bath temperature emerges due to the quantum heating. With increasing decay rates the peaks merge. Superimposed on the shown spectra is the supernarrow peak at frequency $\omega_F/2$ ($\nu=0$ in the rotating frame), which is described by Eq.~(\ref{eq:supernarrow_peak}).}
\label{fig:moderate_damping}
\end{figure}

The evolution of the spectrum (\ref{eq:spectrum_moderate_damping}) with varying oscillator parameters is illustrated in Fig.~\ref{fig:moderate_damping}. With increasing bath temperature the intensity of the spectrum goes up, but the peaks at $\nu_0$ and $-\nu_0$ remain different. The ratio of their intensities approaches $|u/v|^2$ for $\bar n\gg 1$.On the other hand, with decreasing $\nu_a/\kappa$ the peaks start overlapping, and ultimately form a single peak. For small $\nu_a/\kappa$ the peak is centered at $\nu=0$ (at $\omega=\omega_F/2$, in the laboratory frame) and has halfwidth $\nu_a^2/2\kappa \ll \kappa$. The limit $\nu_a/\kappa \ll 1$ is relevant for the vicinity of the bifurcation point $\mu=-(1-\kappa^2)^{1/2}$ where the period-two vibrations disappear.

\section{Supernarrow spectral peak and some generalizations}
\label{sec:supernarrow}

Along with small-amplitude fluctuations around the stable vibrational states, quantum and classical fluctuations lead to occasional interstate switching. Unless the damping rate is extraordinarily small, even for zero bath temperature the switching occurs via transitions over the quasienergy barrier that separates the minima of $g(Q,P)$, see Fig.~\ref{fig:quasienergy} \cite{Marthaler2006}. For small $\lambda(2\bar n+1)$ the transition rate $W_{\rm tr}$ is exponentially small, $W_{\rm tr}\propto \Gamma\exp(-R/\lambda)$ (the dimensional decay rate $\Gamma=\kappa F/2\omega_F$); the effective activation energy $R$ was discussed earlier \cite{Marthaler2006,Dykman2007}.

An important manifestation of interstate switching is the occurrence of an additional peak in the oscillator power spectrum. It is centered at frequency $\omega_F/2$ and is supernarrow in the sense that its width is much smaller than $\Gamma$. The peak is analogous to the supernarrow peak in the spectra of oscillators with coexisting vibrational states in a resonant additive field \cite{Dykman1979a,*Dykman1994b,Stambaugh2006a}. The distinction is that, for a parametrically modulated oscillator, average populations of the stable states are equal for all parameter values in the range of bistability, as a consequence of the symmetry with respect to time translation by $2\pi/\omega_F$. The supernarrow peak in the response of the parametric oscillator was discussed earlier \cite{Ryvkine2006a}, and in the noise spectrum of a parametric oscillator it was recently seen in the experiment \cite{Wilson2010}.

To describe the peak in the noise spectrum we note that the populations $\rho_+$ and $\rho_-$ of the stable vibrational states $(Q_0,P_0)$ and $-(Q_0,P_0)$, respectively, satisfy the balance equation
\begin{equation}
\label{eq:balance_eq}
d\rho_{\pm}/dt = \pm W_{\rm tr}(\rho_- - \rho_+).
\end{equation}
Fluctuations of the populations $\rho_{\pm}$ lead to fluctuations of the expectation values of the operators $a(t), a^{\dagger}(t)$ between the stable-states values $\bigl(a_0(t),a_0^*(t)\bigr)$ and $-\bigl(a_0(t),a_0^*(t)\bigr)$; we note that fluctuations about the stable states are averaged out on time scale $\sim \Gamma^{-1} \ll W_{\rm tr}^{-1}$. The contribution of these fluctuations to the time correlation function of $a,a^{\dagger}$ is
\begin{equation}
\label{eq:fluctuations_a_slow}
\langle\langle a(t)a^{\dagger}(0)\rangle\rangle_{\rm tr} \approx a_0e^{-i\omega_Ft/2}[\rho_+(t;a^{\dagger})-\rho_-(t;a^{\dagger})],
\end{equation}
where $\rho_{\pm}(t;a^{\dagger})$ satisfy Eq.~(\ref{eq:balance_eq}) with initial conditions $\rho_{\pm}(0;a^{\dagger})=\pm a_0^*/2$ that follow from the stationary state populations being equal to 1/2.

From Eqs.~(\ref{eq:spectrum_defined}), (\ref{eq:balance_eq}), and (\ref{eq:fluctuations_a_slow}) we obtain the full expression for the power spectrum as
\begin{equation}
\label{eq:power_complete}
\Phi(\omega)=\Phi_0(\omega) + \Phi_{\rm tr}(\omega),
\end{equation}
where $\Phi_{\rm tr}$ describes the interstate-transition induced contribution,
\begin{equation}
\label{eq:supernarrow_peak}
\Phi_{\rm tr}(\omega)=2W_{\rm tr}|a_0|^2 \left[4W_{\rm tr}^2+ \left(\omega - \frac{1}{2}\omega_F\right)^2\right]^{-1}.
\end{equation}
Function $\Phi_{\rm tr}(\omega)$ has the shape of a Lorentzian peak with halfwidth $2W_{\rm tr} \ll \Gamma$. The intensity of this supernarrow peak is determined by the squared scaled amplitude of the period-two vibrations $\propto P_0^2+Q_0^2$. The area of the peak is independent of the bath temperature, but its width sharply increases with the increasing temperature.

\subsection{Quantum temperature for zero-amplitude states}

In the parameter range $|\mu|> (1-\kappa^2)^{1/2}$ the oscillator has a stable state where the amplitude of vibrations at frequency $\omega_F/2$ is zero. Even though the oscillator does not vibrate on average, fluctuations about the zero-amplitude state are modified by the periodic modulation. These fluctuations are described by Eqs.~(\ref{eq:g}) and (\ref{eq:QKE_operator_form}). For small damping, in the frame rotating at frequency $\omega_F/2$, fluctuations are random vibrations of an auxiliary oscillator at a dimensionless frequency $\nu_0'= (\mu^2-1)^{1/2}$. These vibrations can be described using the Bogoliubov transformation similar to that in Sec.~\ref{sec:quantum_T}, with $a_0=0$ and with $u$ and $v$ replaced by $u'$ and $v'$, respectively,
\begin{eqnarray}
\label{eq:zero-amplitude}
&&u'=-i\left(|1+\mu|^{1/2}+ |1-\mu|^{1/2}\right)/2\nu_0^{\prime\,1/2}, \nonumber\\
&&v'=i\left(|1+\mu|^{1/2}- |1-\mu|^{1/2}\right)/2\nu_0^{\prime\,1/2}.
\end{eqnarray}

The effective Planck number of the vibrations of the auxiliary oscillator is
\begin{eqnarray}
\label{eq:Planck_zero_amplitude}
\bar n_e'= \frac{2\bar n+1}{4\nu_0'}\left(|1+\mu| + |1-\mu|\right)-\frac{1}{2}.
\end{eqnarray}
Even where $\bar n=0$, we have a nonzero  $\bar n_e'=|v'|^2 > 0$. The effective temperature of the auxiliary oscillator increases close to the bifurcation points $\mu\approx \pm 1$ where the zero-amplitude states of the original oscillator loose stability, with $\bar n_e'\approx |\mu^2-1|^{-1/2}/2$ for $\bar n=0$. On the other hand, far from the bifurcation points, where $\mu^2\gg 1$, we have $\bar n_e'\approx \bar n$, i.e., as expected for a zero-amplitude state, the temperature of the auxiliary oscillator approaches the bath temperature.

The distribution over the quasienergy states for the oscillator fluctuating about a zero-amplitude state can be directly measured spectroscopically through the fine structure of the power spectrum. The nonequidistance of the quasienergy levels in this case is the same as for the energy levels of the original oscillator in the absence of modulation.

\section{Conclusions}
\label{sec:conclusions}

This paper was focused on quantum fluctuations that accompany relaxation in modulated oscillators. These fluctuations lead to a finite width of the distribution of the oscillator over quasienergy states, even for zero temperature of the thermal bath that causes relaxation. We call this effect quantum heating. It gives an extra contribution to the standard quantum fluctuations related to a finite width of the oscillator distribution over the coordinate and momentum in each quasienergy state.

As a consequence of the finite width of the quasienergy distribution, the power spectrum of an underdamped oscillator in the rotating frame has peaks at frequencies that correspond to transitions with increasing or decreasing quasienergy. These frequencies have opposite signs, and the ratio of the peak amplitudes is determined by the width of the quasienergy distribution. The transitions occur primarily between neighboring quasienergy levels. We note that, in the language of quantum optics, one can think of the peaks as resulting from parametric down-conversion: a photon at frequency $\omega_F$ splits into photons at frequencies $\omega_F/2 \pm \delta\omega$. However, the processes involved are substantially multiphoton, and a description in terms of quasienergies is more adequate.

We have shown that the peaks of the power spectra may have fine structure. It emerges where the difference in frequencies of transitions between neighboring pairs of quasienergy levels exceeds the decay rate. In dimensionless units this condition has the form $\lambda |V|\gg \kappa$, see Sec.~\ref{sec:fine_structure}. The power spectrum of a nonlinear oscillator may display fine structure also in the absence of periodic modulation, provided the nonequidistance of the energy levels exceeds their width, which is the same condition but applied to the energy rather than quasienergy levels. Quantitatively, it has the form $\lambda \gg 2\kappa$ \cite{Dykman1973}.

For the period-two states $|V| > 2$, and $|V|$ becomes large near the bifurcation point where $\mu+1$ is small, see Eq.~(\ref{eq:g_eigenvalues}). Hence it can be significantly easier to observe the fine structure for a modulated oscillator than for an unmodulated one. In addition, the observation does not require that the excited states of the unmodulated oscillator be thermally populated. A comparatively strong nonequidistance of the energy levels of unmodulated oscillators has been already achieved in circuit QED; in particular, it underlies the operation of the transmon qubits \cite{Schreier2008}. Therefore the fine structure predicted in this paper should be accessible to the experiment. A similar fine structure can be observed in the power spectrum of an oscillator driven by an additive force with frequency close to the oscillator eigenfrequency.

An interesting feature of the parametrically modulated oscillator, that does not occur in an additively driven oscillator, is the occurrence of the parameter value where the effective temperature of the quasienergy distribution coincides with the temperature of the bath, to the leading order in the distance from the stable state along the quasienergy axis. Near bifurcation points where the stable state disappears, on the other hand, the effective temperature sharply increases.

Another important feature is the supernarrow peak at frequency $\omega_F/2$, which emerges in the response \cite{Ryvkine2006a} and also in the noise spectrum, where it has been already seen in the experiment \cite{Wilson2010}. In contrast to the supernarrow peak for additively driven oscillators \cite{Dykman1979a,Stambaugh2006a}, for parametric oscillators the peak has large intensity in a broad parameter range, everywhere where the period-two states are significantly populated. The width of the peak is determined by the rate of switching between the period-two states and is much smaller than the oscillator relaxation rate.

For small damping, $\kappa\ll \lambda |V|$, the quantum-heating induced fine structure should be observable not only in the noise spectrum, but also in the spectrum of linear response to an additional weak field at frequency $\omega$ close to $\omega_F/2 \pm F\nu_0/2\omega_F$. Since near its maximum the quasienergy distribution is of the Boltzmann form, this spectrum can be analyzed using an appropriately modified fluctuation-dissipation relation. Its shape is similar to that described by Eq.~(\ref{eq:partial_general}). Where the fine structure is smeared out, $\kappa\gg \lambda |V|$, quantum effects weakly change the response to an extra weak field, and the analysis of this response for a parametrically modulated oscillator can be done in the same way as for a classical oscillator driven by a resonant additive force \cite{Dykman1979a}. We note that the response emerges both at frequency $\omega$ and the mirror frequency $\omega_F-\omega$.

In conclusion, we have demonstrated that quantum fluctuations, which accompany relaxation of a periodically modulated oscillator, can be observed by studying the oscillator power spectrum. For a parametrically modulated oscillator, we found the spectrum in an explicit form. In the laboratory frame, the spectrum may have two peaks located on the opposite sides of half the modulation frequency, or, for higher damping, a single peak. Where the spectrum has two peaks, the ratio of their intensities is determined by the quantum temperature, which characterizes the distribution over the quasi-energy states of a modulated system.  Generally, it exceeds the bath temperature. For small damping, the spectral peaks may display a fine structure. The intensities of the fine-structure lines as well as their shapes are also determined by, and sensitively depend on the quantum temperature, suggesting an independent way of measuring it.

We gratefully acknowledge the discussion with P. Bertet. The research of MID and VP was supported by the NSF, grant EMT/QIS 082985, and by DARPA DEFYS.

%\bibliographystyle{apsrev}
%\bibliographystyle{apsrev4-1}
%\bibliography{c:/Aaa/Bibtex/md10}

\end{document}